\documentclass[a4paper, 10pt, conference]{IEEEtran}      

\IEEEoverridecommandlockouts                              

\usepackage{amsmath} 
\usepackage{textcomp}
\usepackage{tikz}

\usepackage{siunitx}
\DeclareSIUnit{\pixel}{px}

\makeatletter
\let\labelindent\@undefined 
\makeatother
\usepackage{enumitem}
\usepackage[export]{adjustbox} 

\usepackage{IEEEcopyright}
\usepackage{mynohyperref}
\usepackage{mwe}
\usepackage{subcaption}
\usepackage{cite}

\title{
Improved Calibration Procedure for Wireless Inertial Measurement Units without Precision Equipment
}

\author{
    \IEEEauthorblockN{
        Fritz Webering\IEEEauthorrefmark{1}, 
        Sarah Kleinjohann\IEEEauthorrefmark{2}, 
        Nils Stanislawski\IEEEauthorrefmark{1}, 
        Holger Blume\IEEEauthorrefmark{1} 
    }
    \IEEEauthorblockA{
        \IEEEauthorrefmark{1}Institute of Microelectronic Systems, Leibniz University Hannover, Hannover, Germany
    }
    \IEEEauthorblockA{
        \IEEEauthorrefmark{2}Student at Leibniz University Hannover, Hannover, Germany
    }
    \IEEEauthorblockA{
        \IEEEauthorrefmark{1}\{webering, stanislawski, blume\}@ims.uni-hannover.de
    }
    \IEEEauthorblockA{
        \IEEEauthorrefmark{2}sarah.kleinjohann@stud.uni-hannover.de
    }
}

\def\Vtheta{\boldsymbol{\theta}}
\DeclareSIUnit\gravity{\textit{g}}

\begin{document}

\maketitle
\thispagestyle{empty}
\pagestyle{empty}


\begin{abstract}

Inertial measurement units (IMUs) are used in medical applications for many different purposes. 
However, an IMU's measurement accuracy can degrade over time, entailing re-calibration.
In their 2014 paper, Tedaldi et al. presented an IMU calibration method that does not require external precision equipment or complex procedures.
This allows end-users or personnel without expert knowledge of inertial measurement to re-calibrate the sensors by placing them in several suitable but not precisely defined orientations.
In this work, we present several improvements to Tedaldi's method, both on the algorithmic level and the calibration procedure: adaptions for low noise accelerometers, a calibration helper object, and packet loss compensation for wireless calibration.
We applied the modified calibration procedure to our custom-built IMU platform and verified the consistency of results across multiple calibration runs.
In order to minimize the time needed for re-calibration, we analyzed how the calibration result accuracy degrades when fewer calibration orientations are used.
We found that N=12 different orientations are sufficient to achieve a very good calibration, and more orientations yielded only marginal improvements.
This is a significant improvement compared to the 37 to 50 orientations recommended by Tedaldi.
Thus, we were reduced the time required to calibrate a single IMU from ca. 5 minutes to less than 2 minutes without sacrificing any meaningful calibration accuracy.


\end{abstract}

\section{INTRODUCTION}

Inertial measurement units (IMUs) are used in a wide variety of medical and sports applications, like injury rehabilitation, training progress monitoring, or fall detection in elderly patients, among many others~\cite{Hu2016, ahmad2013reviews, Ganesan2015}.
Even more complex tasks such as full-body human motion capture have also been addressed using inertial sensors~\cite{malleson2017, Zihajehzadeh2017}.
In areas where precise measurements and orientation estimates are required, for example in sports or motion capturing, correct calibration of each unit is crucial.

When the parameters of the embedded inertial sensors change over long times or varying temperatures, the unit will no longer adhere to its specified accuracy limits, and a re-calibration becomes necessary.
The method proposed by Tedaldi et~al.~\cite{Tedaldi2014ARA} in 2014 allows users to re-calibrate their IMUs in the field without the presence of any specialized and often expensive reference equipment like right angles, turntables, or high precision servo platform~\cite{BOTEROVALENCIA2017257}.
When we implemented this method using our in-house wireless IMU platform, which is described in Section~\ref{sec:imu-platform}, we noticed some problems.
After careful examination of the algorithm and associated procedures, we were able to overcome the issues as described in Section~\ref{sec:improvements}.

The calibration method presented by Tedaldi et al. uses raw accelerometer and gyroscope data~\cite{Tedaldi2014ARA} which is processed on a host device.
The original implementation is in C++~\cite{imu_tk}, but for our experiments, we used the MATLAB implementation written by Jianzhu Huai~\cite{imu_tk_matlab} because we found the source code easier to use, understand, and modify.

\section{RELATED WORK}
\label{sec:related-work}

Another calibration scheme without external equipment was proposed by Ren et al.~\cite{Ren2015} in 2015 and is based on the concept of rotating the IMU on an inclined plane, thereby determining the heading using the accelerometer.
The underlying principle is fundamentally similar to the concept of Tedaldi's algorithm, but the calibration protocol is more complex than simply laying the IMU in different random orientations.
While Ren et al. reported exceptional calibration accuracy and mentioned Tedaldi's earlier work, they do not compare the accuracy of the two methods.

Peng et al.~\cite{Peng2022} presented a similar approach to ours in 2022, who also used a 3D-printed icosahedron to orient the IMU during calibration.
They implement a simplified calibration procedure based on an iterative weighted Levenberg–Marquardt algorithm on an embedded microcontroller.
However, they disregard axis misalignment and suppose the availability of an expensive precision multi-axis servo stage for gyroscope calibration.
This makes the algorithm a bit pointless because that same platform could also calibrate the accelerometer.

\section{IMU PLATFORM}
\label{sec:imu-platform}

In order to have a fully configurable IMU development platform, we developed our own wireless sensor device.
The development platform was optimized regarding sensor data accuracy, size, weight, and power consumption.

\begin{figure}[t]
    \centering
    \includegraphics[width=0.73\linewidth,origin=c,valign=t]{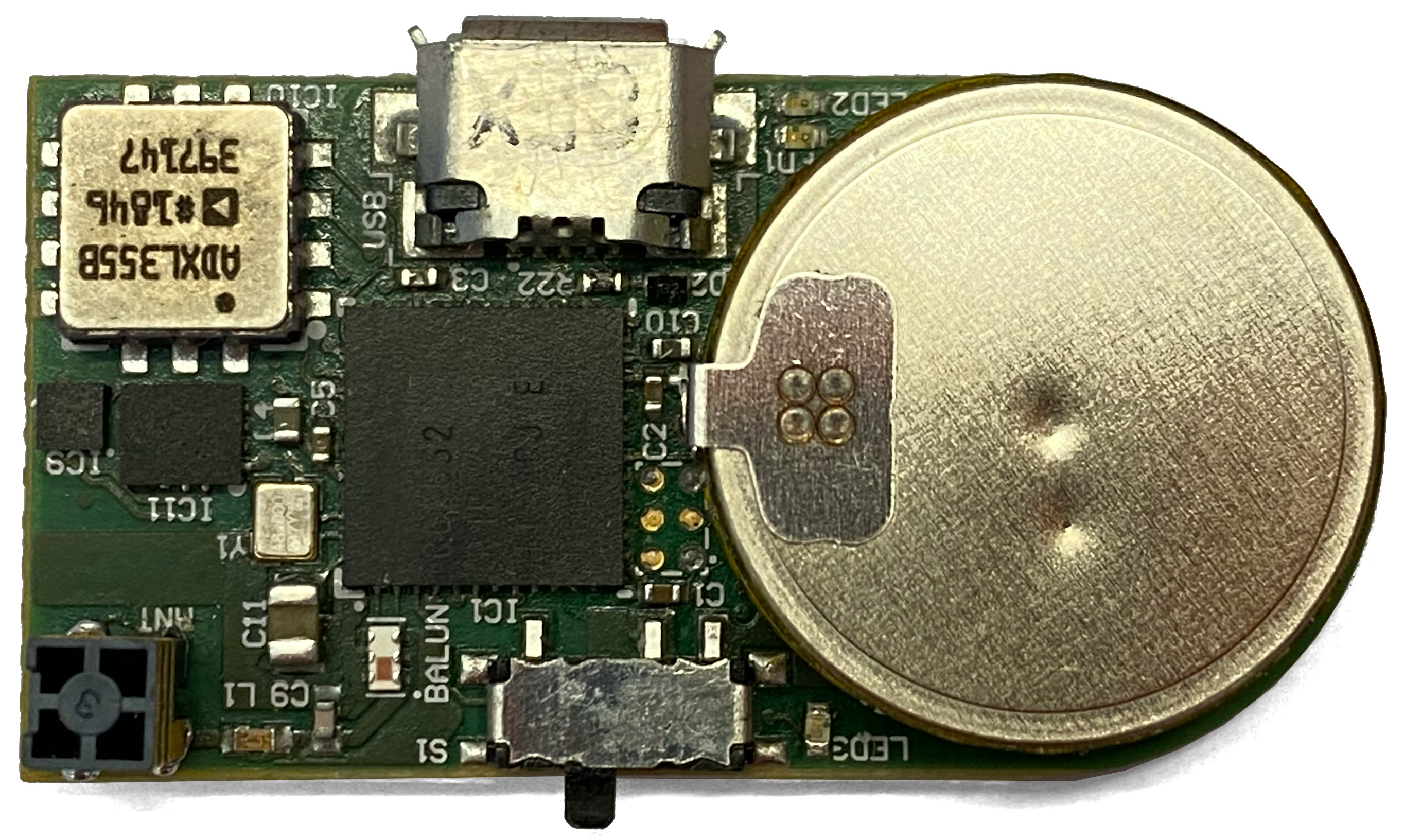}%
    \includegraphics[width=0.27\linewidth,origin=c,valign=t]{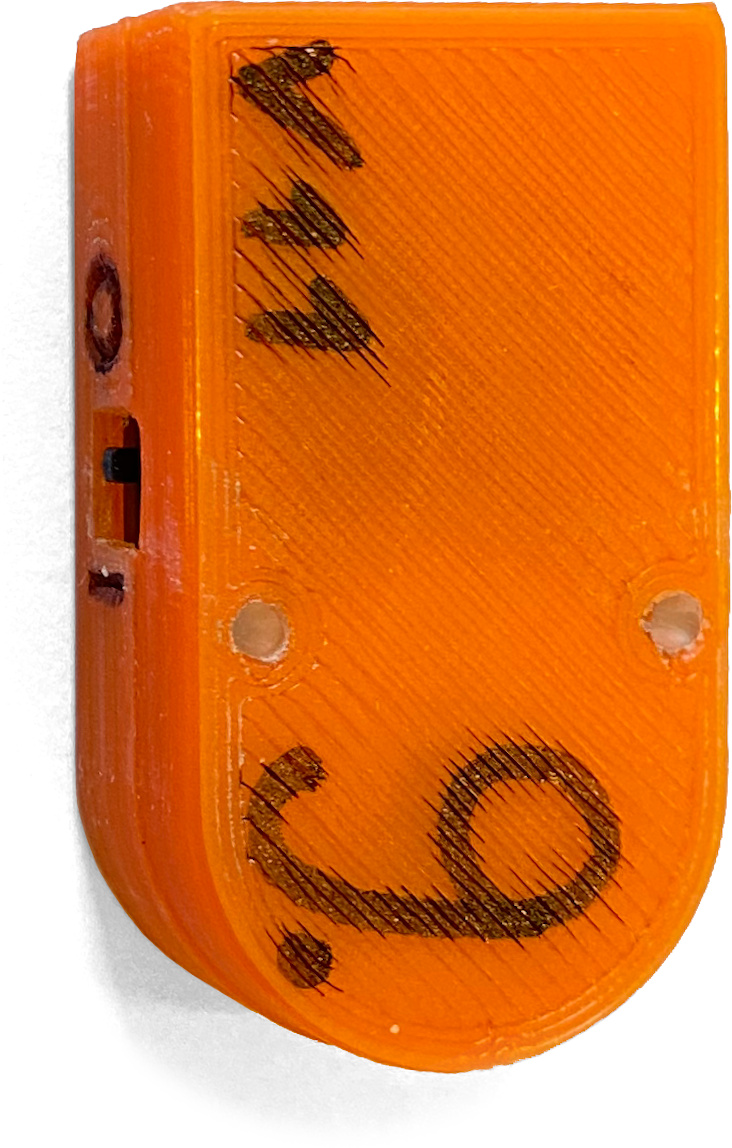}
    \caption{
        Left: PCB of the 6th generation of our in-house developed wireless IMU.
        Right: 3D-printed case, which contains the PCB (IMU \#9, case version 11).
    }
    \label{fig:imu-pcb}
\end{figure}

The System-on-Chip (SoC) CC2652R1 by Texas Instruments was chosen as microcontroller~(µC). 
It includes a 32-bit ARM Cortex M4F core, a 2.4 GHz Bluetooth-Low-Energy~(BLE) transceiver powered by an ARM Cortex M0 core, and a low-power sensor-controller core for interacting with peripheral components without requiring the main core.
The wireless IMU features a single-chip IMU sensor containing both a triaxial accelerometer and gyroscope (BMI160 by Bosch), an additional high-precision accelerometer (ADXL355BEZ by Analog Devices), and a magnetic sensor (MMC3416xPJ by MEMSIC).
The magnetic sensor is used for magnetic, angular rate, and gravity (MARG) sensing applications and enables distortion and gyroscope bias drift compensation.
A barometric pressure sensor~(BMMP388 by Bosch) is included for floor-level detection.
A Serial-to-USB interface (FT230XQ) facilitates wired data transmission charging of the integrated 120\si{\milli\ampere\hour} lithium-ion battery.

Component placement and routing are realized on a four-layer printed circuit board (PCB) with two component sides and dimensions of 34.5\si{\milli\meter} by 18\si{\milli\meter}. 
The populated PCB and its 3D-printed enclosure are depicted in Figure~\ref{fig:imu-pcb}.

All sensors are sampled with a rate of~\SI{100}{\hertz}, and raw sensor data is fused using an algorithm published by Madgwick for either IMU or MARG sensing applications~\cite{madgwick2010efficient}.
Sensor fusion can be performed directly on the microcontroller if desired and reduces the amount of data to be transmitted significantly compared to the transmission of raw sensor data.
This allows for the simultaneous operation of multiple wireless IMUs at a refresh rate of \SI{100}{\hertz}.

\section{CALIBRATION IMPROVEMENTS}
\label{sec:improvements}

\subsection{Orientation Helper Object}

The calibration algorithm requires at least nine different orientations to construct a well-defined optimization problem~\cite{Tedaldi2014ARA,Syed_2007}.
In Tedaldi's original work, the IMU is placed in $37 \leq N \leq 50$ distinct static positions, each held for $1-4$\si{\second}.

Without support, a cuboid IMU case can only be placed on six faces -- or less if some are rounded, as seen in Figure~\ref{fig:imu-pcb}.
In~\cite{Tedaldi2014ARA}, Tedaldi et al. show an IMU with a cable attached resting on the edge of a slab, which is a very unstable position for an IMU weighing only a few grams.
The rigid cable can move the lightweight IMU by tiny amounts, preventing the algorithm from identifying static phases.

\begin{figure}
    \centering
    \includegraphics[width=0.5\linewidth]{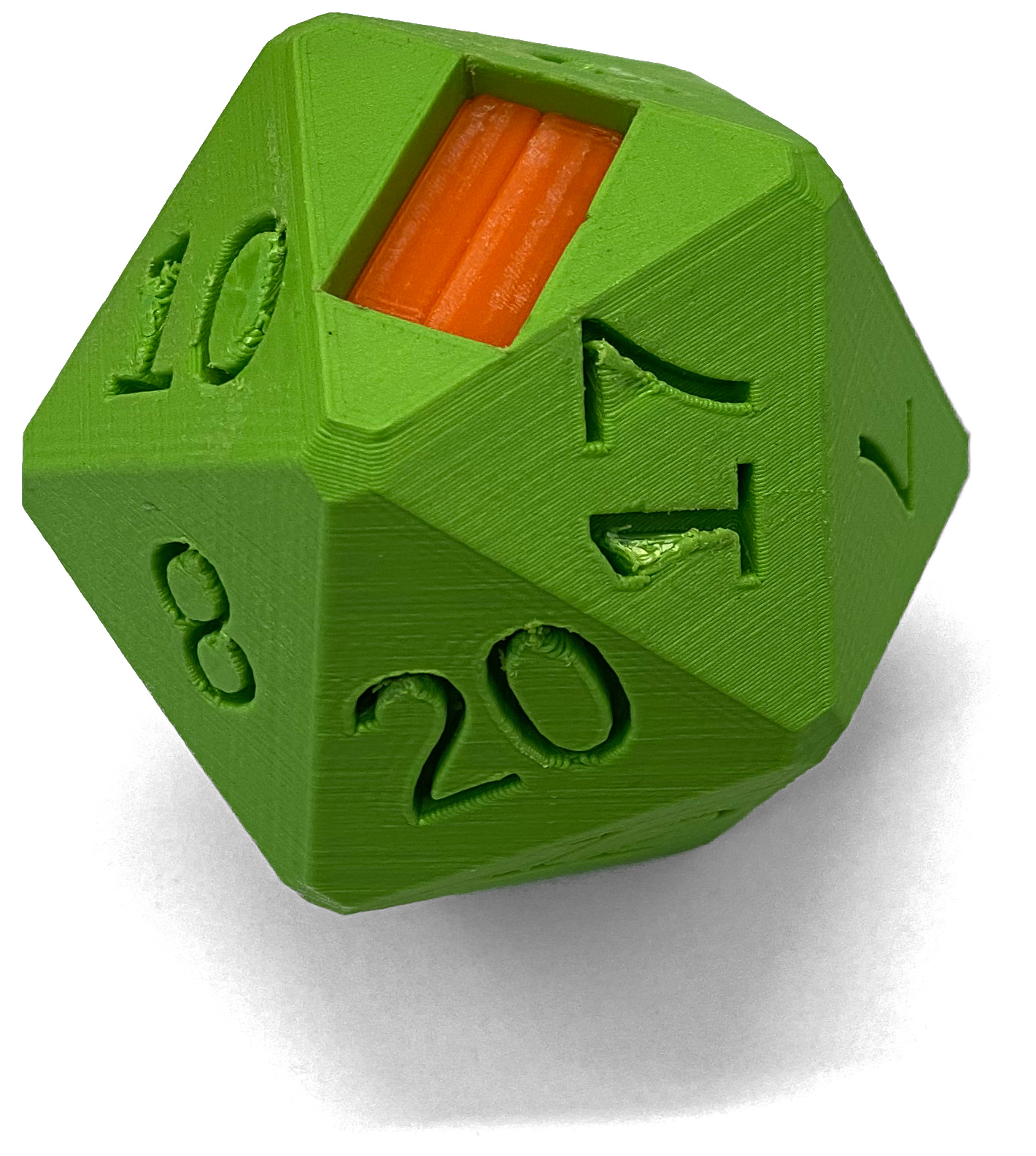}%
    \includegraphics[width=0.5\linewidth]{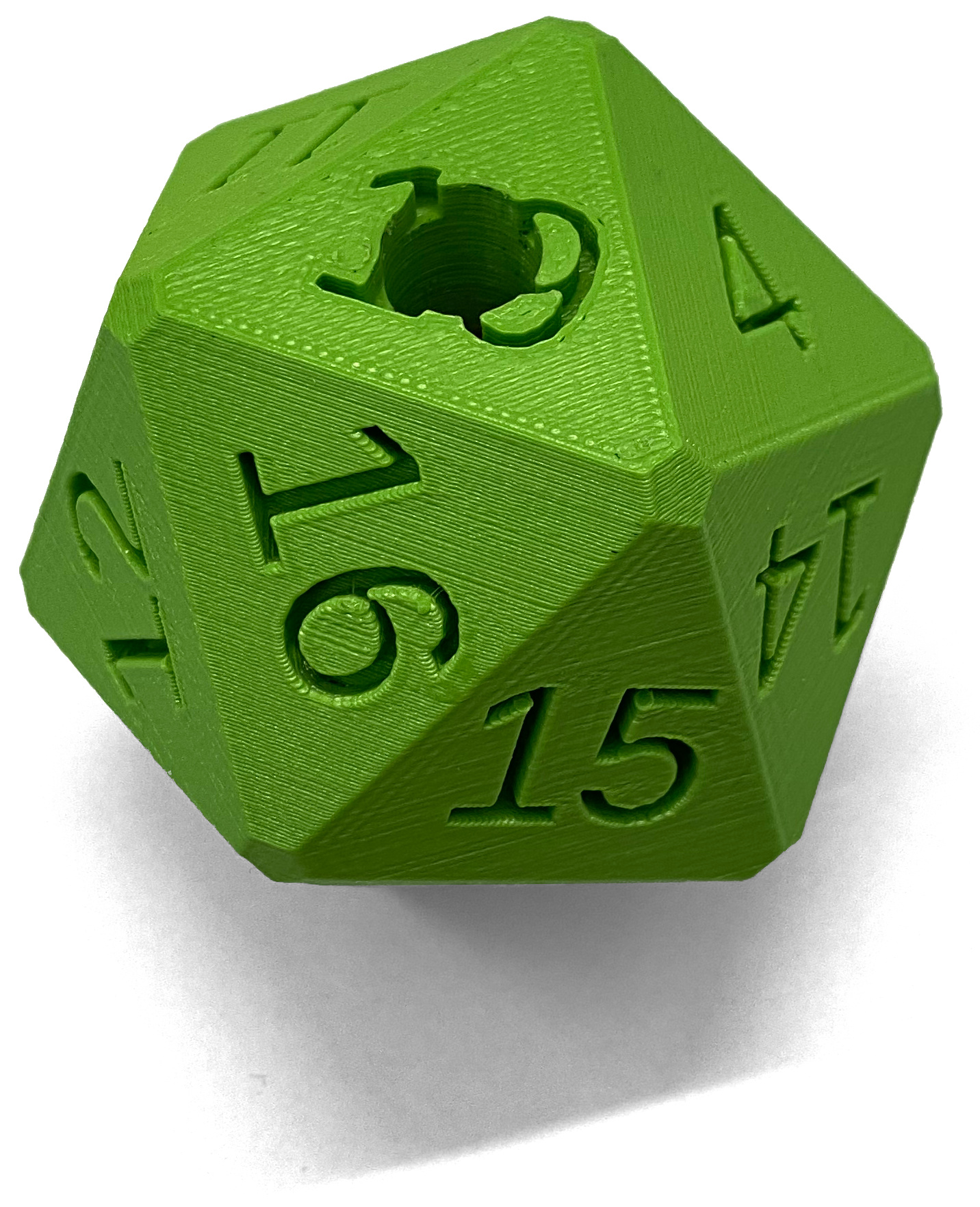}
    \caption{The icosahedral enclosure allows the IMU to rest in 20 different orientations. It can hold the IMU case (orange) and has a hole opposite for pushing the IMU back out.}
    \label{fig:icosahedron}
\end{figure}

To avoid these problems in our evaluation, we use wireless transmission (see also Section~\ref{sec:packet-loss}) and a 3D-printed icosahedral orientation helper object based on Tim Edwards' OpenSCAD model~\cite{thingiverse-dice} as shown in Figure~\ref{fig:icosahedron}.
The 3D-printed regular icosahedron has a distance of \SI{66}{\milli\meter} between opposing vertices.
The enclosed IMU is press-fit into a slot on one of the triangular faces and can be released by pushing through a  hole on the opposing face.
The orientation helper object allowed us to place the IMUs to be calibrated in 20 easily reproducible orientations with at least 42 degrees rotation between them and without interfering cables.
This enabled a more systematic approach to capturing calibration sequences, as explained in Section~\ref{sec:evaluation}.

\subsection{Improved Static Phase Selection}

Running the calibration algorithm~\cite{imu_tk_matlab} on our captured sequences of the low noise accelerometer revealed a weakness of the static phase detection algorithm.
This issue was found in both the original C++ code~\cite{imu_tk} and the MATLAB implementation~\cite{imu_tk_matlab}.
The variance of the long static phase in the beginning $\varsigma_\text{init}$ is used as a baseline for finding a suitable variance threshold for the short static segments later in the sequence.
However, $\varsigma_\text{init}$ is so small due to the low noise floor of the ADXL355 that even tiny perturbations in the short sequences are above the maximum threshold of $10 \varsigma_\text{init}$ variance.
Increasing the maximum variance threshold factor from 10 to 225 solved this problem for our case but revealed another problem:
v
In Tedaldi's algorithm, the best static phase threshold (an integer multiple $k \cdot \varsigma_\text{init}$) is determined by performing a non-linear least-squares minimization of the accelerometer cost function $\boldsymbol L(\Vtheta^\text{acc})$ for each $k$ and selecting the $k$ with the smallest residual.
As stated in~\cite{Tedaldi2014ARA}, this approach does not require parametrization, but favors calibration sequences in which the same orientation is repeated multiple times.
Thus, the algorithm would select a $k$ for which the long sequence at the beginning was split into multiple smaller segments of length $> \SI{1}{\second}$ because the variance $\varsigma(t)$ was too close to $k \varsigma_\text{init}$, crossing the threshold multiple times.
This problem can be fixed by rejecting segments where the acceleration vector direction did not change relative to the previously accepted segment, minimizing the number of redundant segments.

However, the selection of the $k$ with the minimum residual was still susceptible to the slightest perturbation of the accelerometer data.
When truncating the calibration sequence even slightly, the algorithm would seemingly at random select very different values of $k$, which led to wildly varying numbers of segments:
Fewer segments for smaller $k$ because static phases would be broken into multiple parts, which were then rejected for being shorter than \SI{1}{\second}.
Thus, we modified the static phase selection to always select the $k$ which produced the largest number of usable static segments, excluding segments that were duplicates or too short.
If there are multiple $k$, we select the one with the lowest residual.
\footnote{Our changes to the MATLAB code, including evaluation scripts, will be published under \texttt{\scriptsize https://github.com/IMS-AS-LUH/imu\_tk\_matlab}}

\subsection{Division by zero}

The MATLAB implementation~\cite{imu_tk_matlab} contained a division by zero in the function \texttt{fromOmegaToQ} when the angular rates were $[0,0,0]$ in a single packet.
This is a very unlikely event due to the measurement noise, but it occurred in at least one calibration sequence, so we corrected the issue.

\subsection{Wireless Packet Loss Correction}
\label{sec:packet-loss}

For data transmission between IMU and host PC, we used Bluetooth 5.0 Low Energy~(LE) Generic ATTribute Profile~(GATT) notifications since they provide a significant improvement in data transmission speed compared to indication messages~\cite{blespec}.
Even though the Bluetooth link layer L2CAP provides acknowledgments and retransmissions, any wireless transmission includes a risk of packet loss.
Depending on the application, the loss of a single quaternion may not be critical, but the loss of raw gyroscope data in a calibration sequence will result in integration errors.
Loss of accelerometer data is not critical because it is only sampled in static periods.

Thus, we implemented simple and a power-efficient erasure code~(EC) for forward error correction of lost gyroscope samples in order to prevent this problem.
The EC data $E_i$ is the same size as the raw gyroscope data $_i$ (6 bytes) and consists of the XOR of the previous $M$ raw gyroscope values $E_i = G_{i-1} \bigoplus G_{i-2} \bigoplus \ldots  \bigoplus G_{i-M}$. 
This allows the receiver to reconstruct arbitrary packet losses in a window of length $M$ when followed by a sequence of $M$ correctly received packets.
Computing $E_i$ for a new packet consists of only two XOR operations:
One for adding the next $G_{i-1}$ and one for removing the oldest element $G_{i-M}$ from the running XOR sum.
Apart from the running sum $E_i$, only an additional ring buffer for storing $G_i$...$G_{i-M+1}$ is required on the IMU.

\section{EVALUATION}
\label{sec:evaluation}

\begin{figure}[b]
    \centering
    \includegraphics[width=0.9\linewidth]{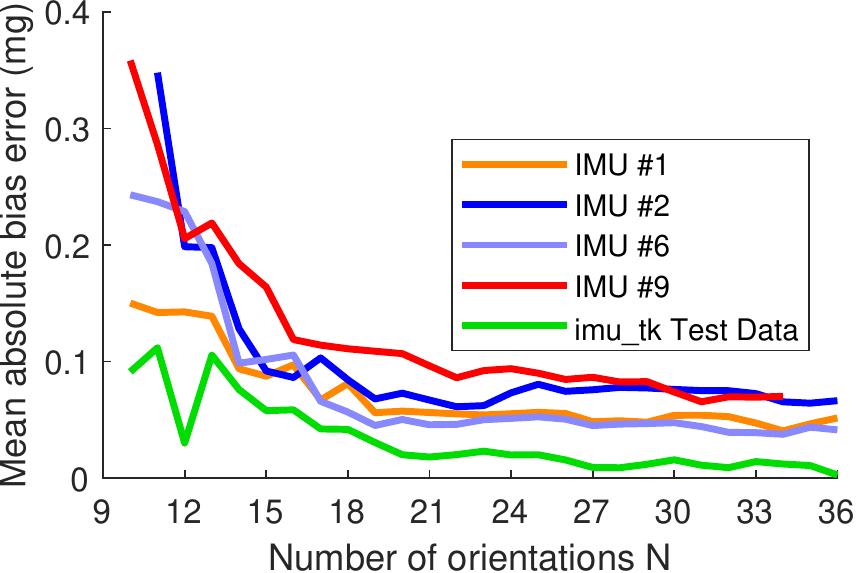}
    \caption{
        Mean absolute difference of accelerometer bias estimation to $\Vtheta_\text{mean}$ for varying $N$. 
        Mean of 5 calibration runs for each IMU except imu\_tk Test Data (only 1 calibration run).
        Units are \SI{1}{\milli\gravity} = \SI{9.807e-3}{\meter\per\second^2}
    }.
    \label{fig:accel-bias}
\end{figure}

To determine the minimum required number of static positions $N$ for a sufficiently accurate calibration, we recorded five calibration sequences with each of our four working IMUs (units \#1, \#2, \#6 and \#9)  with $N \geq 37$.
For each sequence, we recorded the following raw IMU data: Packet index, ADXL355 acceleration, BMI160 acceleration, and BMI160 gyroscope values.
The initial static phase was \SI{40}{\second} long, followed by static segments of $\geq\SI{3}{\second}$.
For the first 20 orientations, we placed the icosahedron on all 20 faces, with increasing numbers pointing up.
After the ascending sequence, we placed the icosahedron on random faces until at least 37 poses were recorded.
One calibration run of IMU~\#9 contained only 34 usable poses, so we discarded all results for IMU~\#9 for $N>34$.
For each IMU, we performed a full calibration for each captured sequence with maximum $N$ using our improved MATLAB code, resulting in 4 times five sets of 18 calibration parameters $\Vtheta = \left[\Vtheta^\text{acc}, \Vtheta^\text{gyro} \right]$, as described in~\cite{Tedaldi2014ARA}).
For each IMU, we calculated the mean parameters $\Vtheta_\text{mean}$ over the five sets, which were subsequently regarded as the `reference' calibration coefficients for that IMU.

\begin{figure*}[!ht]
    \centering
    \begin{subfigure}{.44\linewidth}
        \includegraphics[width=\linewidth]{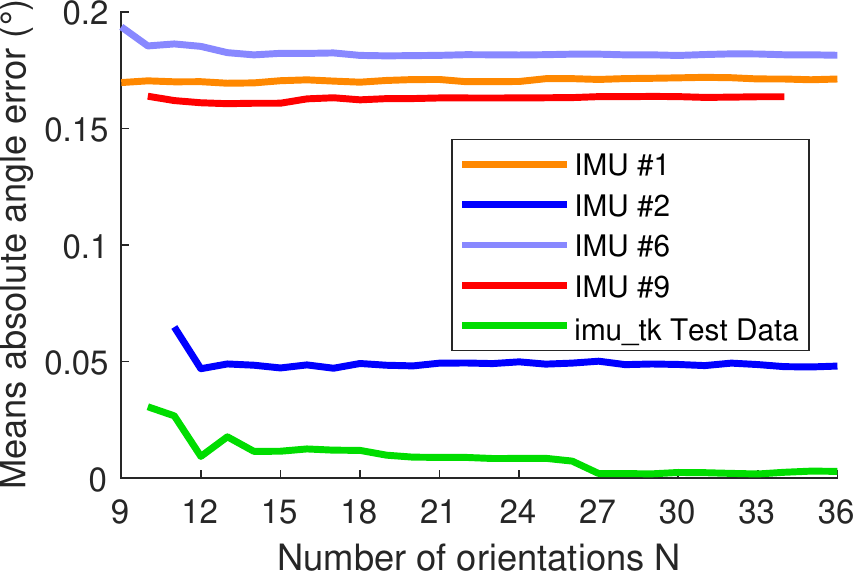}%
        \caption{Gyroscope axis misalignment estimation error over $N$}%
        \label{fig:axis-misalignment:gyro}%
    \end{subfigure}\hfill%
    \begin{subfigure}{.44\linewidth}
        \includegraphics[width=\linewidth]{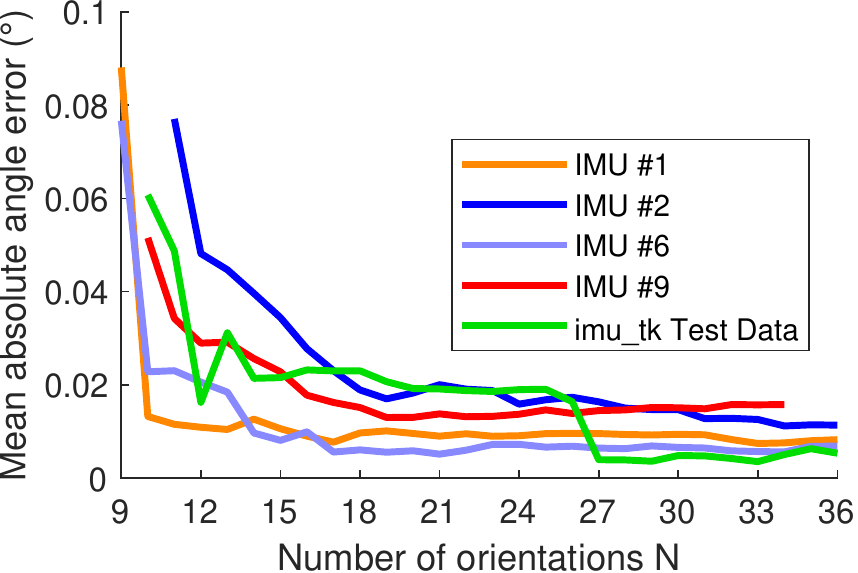}%
        \caption{Accelerometer axis misalignment estimation error over $N$}%
        \label{fig:axis-misalignment:accel}%
    \end{subfigure}\hfill%
    \caption{Axis misalignment estimation errors (including non-orthogonality) for different $N$, when compared to $\Vtheta_\text{mean}$. Typical mean absolute misalignment angles of $\Vtheta_\text{mean}$ for our IMUs were in the range between 0.4 and 0.6 degrees.}
    \label{fig:axis-misalignment}
\end{figure*}

\begin{figure*}[!ht]
    \centering
    \begin{subfigure}{.44\linewidth}
        \includegraphics[width=\columnwidth]{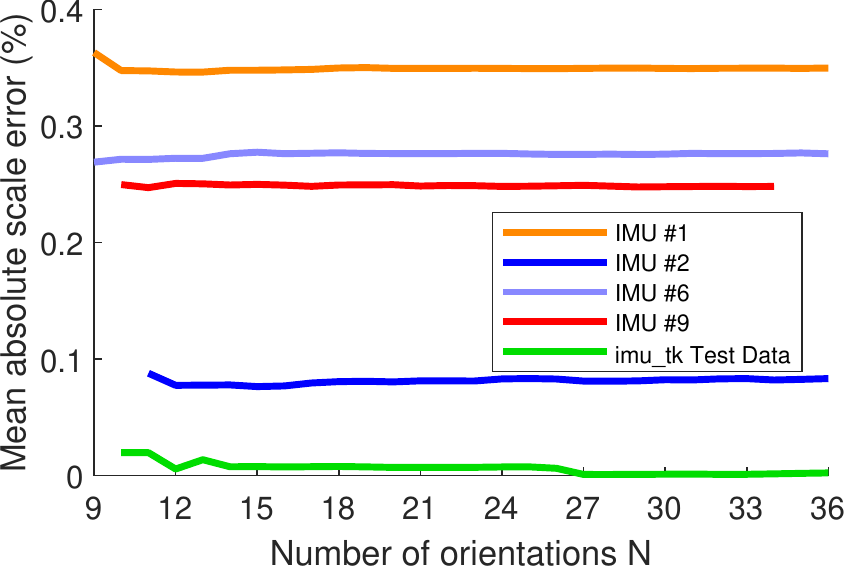}%
        \caption{Gyroscope scaling factor estimation error over $N$}%
        \label{fig:scaling-error:gyro}%
    \end{subfigure}\hfill%
    \begin{subfigure}{.44\linewidth}
        \includegraphics[width=\columnwidth]{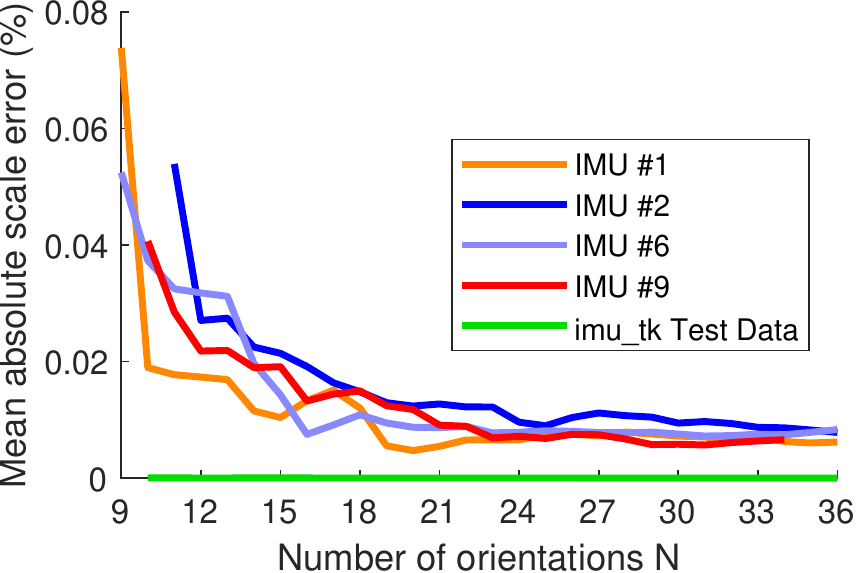}%
        \caption{Accelerometer scaling factor estimation error over $N$}%
        \label{fig:scaling-error:accel}%
    \end{subfigure}\hfill%
    \caption{Scaling factor estimation errors for different numbers of orientations $N$, when compared to $\Vtheta_\text{mean}$. Values are given in percent of the measured quantity (angular rate and acceleration respectively).}
    \label{fig:scaling-error}
\end{figure*}

We then incrementally truncated each individual sequence, thereby reducing the effective $N = N_\text{eff}$.
For each truncated sequence, we performed another calibration run with $N_\text{eff}$ segments and compared the resulting $\Vtheta_{N_\text{eff}}$ to $\Vtheta_\text{mean}$ by calculating the mean absolute difference for each subset of coefficients.
This allows an assessment of how the calibration quality deteriorates with decreasing $N$, relative to the `full' calibration recommended by Tedaldi, without needing an absolute `gold standard' reference of the calibration parameters.

We also captured an additional set of calibration runs without the icosahedron just by placing the IMUs on the five orthogonal faces of their case in four different horizontal attitudes.
The calibration using these sequences completed successfully, but the calibration error for these runs was much larger than $\Vtheta_\text{mean}$, so we omitted the results.

\section{RESULTS}
\label{sec:results}

The results of the evaluation are shown in Figures~\ref{fig:accel-bias}, \ref{fig:axis-misalignment} and \ref{fig:scaling-error}.
Since the conclusion is the same for the BMI160 accelerometer, we only show the results from the calibration runs using the ADXL355 values.
For the accelerometer calibration, we see a slight decrease in error with larger $N$.
However, the errors around $N=12$ orientations are already less than \SI{0.1}{\percent} of the reference values from $\Vtheta_\text{mean}$ for scaling and misalignment.
For bias, the calibration error is already close to the noise floor of the ADXL355 (\SI{200}{\micro\gravity} at \SI{100}{\hertz}). 
Thus, it is questionable whether the additional effort for triple the orientations is justified.

For the gyroscopes, the results are even clearer:
As soon as enough orientations have been captured to compute a successful calibration, the accuracy barely increases at all when more positions are considered.

The results for imu\_tk Test Data are unrealistically low because $\Vtheta_\text{mean}$ is calculated from only one calibration run instead of five, so the calibration sequence is essentially compared to itself. Despite this, the main conclusion for $N$ holds for this IMU as well.


\section{CONCLUSIONS AND OUTLOOK}
\label{sec:conclusion}

In this paper, we presented improvements to a calibration procedure for wireless IMUs which does not require expensive calibration equipment.
The procedures presented in the initial publication by Tedaldi et al.~\cite{Tedaldi2014ARA} and its MATLAB implementation\cite{imu_tk_matlab} were improved upon by correcting errors occurring in rare instances and enhancing the static phase detection when using low-noise accelerometers.
A low-cost and low-accuracy 3D-printed icosahedron served as the only additional and optional calibration equipment and enabled easy positioning of the IMU in 20 distinct attitudes, although our results show that a dodecahedron would probably suffice.
Evaluation of the calibration procedure showed that positioning the IMU in $N=12$ distinct position already results in a minimal error in for the gyroscope calibration and a very small error for the accelerometer $N$ as recommended by Tedaldi et al.~\cite{Tedaldi2014ARA}.

In the next step, it is planned to implement the algorithm presented by Tedaldi et al. directly on the CC2652R1 SoC of our in-house wireless IMU using an approach similar to Peng et al.\cite{Peng2022}, but without sacrificing gyroscope and axis misalignment calibration.
For this purpose, the algorithm's structure will need to be completely remodeled, optimizing for computational complexity, program size, and RAM usage.



\clearpage

\bibliography{bibliography.bib,manual.bib}

\begin{thebibliography}{10}
\providecommand{\url}[1]{#1}
\csname url@samestyle\endcsname
\providecommand{\newblock}{\relax}
\providecommand{\bibinfo}[2]{#2}
\providecommand{\BIBentrySTDinterwordspacing}{\spaceskip=0pt\relax}
\providecommand{\BIBentryALTinterwordstretchfactor}{4}
\providecommand{\BIBentryALTinterwordspacing}{\spaceskip=\fontdimen2\font plus
\BIBentryALTinterwordstretchfactor\fontdimen3\font minus
  \fontdimen4\font\relax}
\providecommand{\BIBforeignlanguage}[2]{{%
\expandafter\ifx\csname l@#1\endcsname\relax
\typeout{** WARNING: IEEEtran.bst: No hyphenation pattern has been}%
\typeout{** loaded for the language `#1'. Using the pattern for}%
\typeout{** the default language instead.}%
\else
\language=\csname l@#1\endcsname
\fi
#2}}
\providecommand{\BIBdecl}{\relax}
\BIBdecl

\bibitem{Hu2016}
\BIBentryALTinterwordspacing
X.~Hu and X.~Qu, ``Pre-impact fall detection,'' \emph{{BioMedical} Engineering
  {OnLine}}, vol.~15, no.~1, Jun. 2016. [Online]. Available:
  \url{https://doi.org/10.1186/s12938-016-0194-x}
\BIBentrySTDinterwordspacing

\bibitem{ahmad2013reviews}
N.~Ahmad, R.~A.~R. Ghazilla, N.~M. Khairi, and V.~Kasi, ``Reviews on various
  inertial measurement unit (imu) sensor applications,'' \emph{International
  Journal of Signal Processing Systems}, vol.~1, no.~2, pp. 256--262, 2013.

\bibitem{Ganesan2015}
\BIBentryALTinterwordspacing
Y.~Ganesan, S.~Gobee, and V.~Durairajah, ``Development of an upper limb
  exoskeleton for rehabilitation with feedback from {EMG} and {IMU} sensor,''
  \emph{Procedia Computer Science}, vol.~76, pp. 53--59, 2015. [Online].
  Available: \url{https://doi.org/10.1016/j.procs.2015.12.275}
\BIBentrySTDinterwordspacing

\bibitem{malleson2017}
C.~Malleson, A.~Gilbert, M.~Trumble, J.~Collomosse, A.~Hilton, and M.~Volino,
  ``Real-time full-body motion capture from video and imus,'' in \emph{2017
  International Conference on 3D Vision (3DV)}, 2017, pp. 449--457.

\bibitem{Zihajehzadeh2017}
S.~Zihajehzadeh and E.~J. Park, ``A novel biomechanical model-aided imu/uwb
  fusion for magnetometer-free lower body motion capture,'' \emph{IEEE
  Transactions on Systems, Man, and Cybernetics: Systems}, vol.~47, no.~6, pp.
  927--938, 2017.

\bibitem{Tedaldi2014ARA}
D.~Tedaldi, A.~Pretto, and E.~Menegatti, ``A robust and easy to implement
  method for imu calibration without external equipments,'' \emph{2014 IEEE
  International Conference on Robotics and Automation (ICRA)}, pp. 3042--3049,
  2014.

\bibitem{BOTEROVALENCIA2017257}
\BIBentryALTinterwordspacing
J.~Botero-Valencia, D.~Marquez-Viloria, L.~Castano-Londono, and
  L.~Morantes-Guzmán, ``A low-cost platform based on a robotic arm for
  parameters estimation of inertial measurement units,'' \emph{Measurement},
  vol. 110, pp. 257--262, 2017. [Online]. Available:
  \url{https://www.sciencedirect.com/science/article/pii/S0263224117304360}
\BIBentrySTDinterwordspacing

\bibitem{imu_tk}
\BIBentryALTinterwordspacing
A.~Pretto, ``imu\_tk c++ implementation,'' 2014. [Online]. Available:
  \url{https://bitbucket.org/alberto_pretto/imu_tk/}
\BIBentrySTDinterwordspacing

\bibitem{imu_tk_matlab}
\BIBentryALTinterwordspacing
J.~Huai, ``imu\_tk\_matlab implementation,'' 2017. [Online]. Available:
  \url{https://github.com/JzHuai0108/imu_tk_matlab}
\BIBentrySTDinterwordspacing

\bibitem{Ren2015}
C.~Ren, Q.~Liu, and T.~Fu, ``A novel self-calibration method for mimu,''
  \emph{IEEE Sensors Journal}, vol.~15, no.~10, pp. 5416--5422, 2015.

\bibitem{Peng2022}
C.-C. Peng, J.-J. Huang, and H.-Y. Lee, ``Design of an embedded icosahedron
  mechatronics for robust iterative imu calibration,'' \emph{IEEE/ASME
  Transactions on Mechatronics}, vol.~27, no.~3, pp. 1467--1477, 2022.

\bibitem{madgwick2010efficient}
S.~Madgwick \emph{et~al.}, ``An efficient orientation filter for inertial and
  inertial/magnetic sensor arrays,'' \emph{Report x-io and University of
  Bristol (UK)}, vol.~25, pp. 113--118, 2010.

\bibitem{Syed_2007}
\BIBentryALTinterwordspacing
Z.~F. Syed, P.~Aggarwal, C.~Goodall, X.~Niu, and N.~El-Sheimy, ``A new
  multi-position calibration method for {MEMS} inertial navigation systems,''
  \emph{Measurement Science and Technology}, vol.~18, no.~7, pp. 1897--1907,
  may 2007. [Online]. Available:
  \url{https://doi.org/10.1088/0957-0233/18/7/016}
\BIBentrySTDinterwordspacing

\bibitem{thingiverse-dice}
\BIBentryALTinterwordspacing
T.~Edwards, ``Openscad polyhedral dice,'' 2015. [Online]. Available:
  \url{https://www.thingiverse.com/thing:1043661}
\BIBentrySTDinterwordspacing

\bibitem{blespec}
\emph{Generic Attribute Profile (GATT)}, Bluetooth SIG, 1 2022, rev.
  GATT.TS.p21.

\end{thebibliography}
\bibliographystyle{IEEEtran}

\end{document}